\documentclass[12pt]{article}%

\newcommand{\eq}[1]{(\ref{#1})}
\newcommand{\eqs}[2]{(\ref{#1},~\ref{#2})}

\newcommand{\Eq}{Eq.~\eq}
\newcommand{\Eqs}{Eqs.~\eqs}

\usepackage{amsmath}%

\begin{document}

\title{On gravitational-electromagnetic resonance}
\author{Michael B. Mensky
\\P.N.Lebedev Physical Institute, \\
53 Leninsky prosp., 119991 Moscow, Russia }

\date{}

\maketitle

\begin{abstract}
This is an English translation of the paper M.B.Mensky, in: K.P.Sta\-nyu\-ko\-vich (ed.), \textit{Problems of Theory of Gravity and Elementary Particles}, issue 6, Moscow, Atomizdat, 1975, p.181-190 (in Russian). This paper elaborates further the idea (formulated in 1971 by Braginsky and Mensky) of detecting high-frequency gravitational waves by observing resonance action of a gravitational wave on the electromagnetic wave in a closed resonator (waveguide). The phenomenon underlying such a detector was called gravitational-electromagnetic resonance (GER). In the present paper both closed (for example circular) resonator or waveguide and long (for example in the shape of a spiral) waveguide are considered as possible gravitational-wave detectors. High-frequency gravitational-wave detectors are now again actual (see A.M.Cruise and R.M.J.Ingley, Class. Quant. Grav. \textbf{22}, S479, 2005), but the current literature on this topic does not cover all the issues discussed in the present paper. 
\end{abstract}

\newpage

The principle of detecting gravitational waves based on the resonance action of the gravitational wave on an electromagnetic wave propagating in a waveguide or resonator has been proposed in the papers \cite{1}. This result was later discussed in the papers \cite{2,3} where other approaches to the consideration of the process and other constructions for realizing the gravitational-electromagnetic resonance (GER) were considered. Below a theoretical model is presented which allows one to rather simply perform a sufficiently complete theoretical analysis of the problem. Some new variants of the devices for realizing GER are considered in the framework of this model. 

\section{Waveguide in the field of a gravitational wave}
\label{WaveguideInGravWave}

The weak polarized gravitational wave propagating in the direction of the axis $x^3$ may be presented by the metric with the following nonzero components \cite{4}: 
\begin{equation}
	g_{00}=-g_{33}=1, \quad g_{11}=-1+a, \quad g_{22}=-1-a, \quad g_{12}=g_{21}=b
\label{1}\end{equation}
where $a$ and $b$ are functions of $x^0-x^3$ which are small comparing to unity. An infinitely thin electromagnetic waveguide immersed in this gravitational field may be described as a two-dimensional surface $\Sigma$, the world surface of this waveguide, embedded in the four-dimensional space-time. In case of a high frequency gravitational wave each point of the waveguide can be considered to move along a geodesic line. The lines $\mathbf{x}=\mathrm{const}$ are geodesic in the first order in $a$, $b$, so that the surface $\Sigma$ may be parameterized as follows: 
\begin{equation}
	x^0=\tau, \quad \mathbf{x}=\mathbf{x}(l).
\label{2}\end{equation}
The parameter $l$ labels the points of the waveguide while $\tau$ is a canonical parameter (proper time) on the trajectory (which is a geodesic line) of each of these points. The derivatives 
\begin{equation}
	n_i(l)=\frac{dx^i}{dl}, \quad i=1,2,3,
\label{3}\end{equation}
are components of the tangent vector to the line describing the shape of the waveguide. We shall choose the parameter $l$ so that this tangent vector have unite length, $\mathbf{n}^2=1$. 

The parameters $(\tau, l)=(\xi^0,\xi^1)$ may serve as coordinates on the surface $\Sigma$, and the metric on this surface is induced by the metric in the whole space-time according to the formula 
\begin{equation}
	G_{ab}(\xi)=g_{\mu\nu}(x(\xi))\cdot \frac{\partial x^\mu}{\partial\xi^a} \frac{\partial x^\nu}{\partial\xi^b} 
	\qquad (a,b=0,1).
\label{4}\end{equation}
The function $x(\xi)$ gives a parametric description of the surface $\Sigma$. Making use of \Eqs{2}{3}, we can show that this induced metric has the form 
\begin{equation}
	G_{00}=1, \quad G_{11}=-1+aH+bF, \quad G_{12}=G_{21}=0
\label{5}\end{equation}
where it is denoted
\begin{equation}
	H(l)=n^{2}_{1}-n^{2}_{2}, \quad F(l)=2n_1n_2.
\label{6}\end{equation}

Assume for simplicity that the group and phase velocities in the waveguide are equal to each other and to the velocity of light $c$. Then the propagation of an electromagnetic way in the waveguide may be considered as the motion of the wave in the two-dimensional Riemannian space $\Sigma$ with metric $G$. In the approximation of geometric optic and with polarization neglected, the wave in the wave-guide is presented by its phase $\varphi(\xi)$ satisfying the eikonal equation \cite{4}
\begin{equation}
	G^{ab}\, \frac{\partial\varphi}{\partial\xi^a} \; \frac{\partial\varphi}{\partial\xi^b}=0.
\label{7}\end{equation}
Making use of Eq.~(\ref{5}) for the metric tensor $G_{ab}$, going over to $G^{ab}$ and factorizing the resulting form of \Eq{7}, we reduce it to the following one: 
\begin{equation}
	\left[\frac{\partial\varphi}{\partial\tau} 
	+\left(1+\frac{aH}{2}+\frac{bF}{2}\right)\frac{\partial\varphi}{\partial l}\right]
	\cdot
	\left[\frac{\partial\varphi}{\partial\tau} 
	-\left(1+\frac{aH}{2}+\frac{bF}{2}\right)\frac{\partial\varphi}{\partial l}\right]
	=0.
\label{8}\end{equation}

The solution to this equation which annulates one of the factors describes a running (in one of two directions) wave. Consider the wave running in the positive direction of the axis $l$ i.e. satisfying the following equation: 
\begin{equation}
	\frac{\partial\varphi}{\partial\tau} 
	+\left(1+\frac{aH}{2}+\frac{bF}{2}\right)\frac{\partial\varphi}{\partial l}=0. 
\label{9}\end{equation}
Solving this equation by iterations and taking $\varphi=k\cdot(\tau - l)$ (with $k=\omega/c=\mathrm{const}$) as zero approximation, we obtain in the first approximation
\begin{align}
	\varphi(\tau, l)=k\cdot(\tau-l)+\chi(\tau, l), \label{10}\\
	\frac{\partial\chi}{\partial\tau}+\frac{\partial\chi}{\partial l}
	=\frac{k}{2}(aH+bF). \label{11}
\end{align}

This equation may be easily solved after transition to the variables $\lambda=\tau+l$ and $\nu=\tau-l$. We have then 
\begin{equation}
	\chi(\tau, l)
	=\frac{k}{4}\int^{\tau+l}_{0}d\lambda
	\left[a(\tau'-x^3(l'))H(l')
	+b(\tau'-x^3(l'))F(l')\right]
	+f(\tau-l),
\label{12}\end{equation}
where it is denoted 
$$
	\tau'= \frac{\lambda+\tau-l}{2} , \quad l'=\frac{\lambda-\tau+l}{2} 
$$
and $f$ is an arbitrary function of a single argument. 

This formula will serve as a basis for the further analysis. However, it is convenient to go over (by the rotation in the plane $x^1$,~$x^2$) to the coordinate system in which $b\equiv 0$. Besides, considering only a single Fourier component of the gravitational wave, we shall take 
\begin{equation}
	a(\tau)=a_0\cos k_g\tau
\label{13}\end{equation}
where $k_g=\omega_g/c$ is a wave number of the gravitational wave. Then 
\begin{align}
		\chi(\tau, l)
		& =\frac{ka_0}{4}\int^{\tau+l}_{0}d\lambda\;
		H\left(\frac{\lambda-\tau+l}{2}\right)
		\cos k_g\left[\frac{\lambda+\tau-l}{2}-x^3\left(\frac{\lambda-\tau+l}{2}\right)\right] \nonumber\\
		& +f(\tau-l)
\label{14}\end{align}

Consider separately the two essentially different situations: 1)~closed waveguide in which the wave propagates freely and 2)~long waveguide with a harmonic signal in the input. 

\section{Closed waveguide}
\label{sec:ClosedWaveguide}

Let the waveguide be closed and has length $L$. Assume at the moment that it lies in the plane $x^3=0$. Then Eq.~(\ref{14}) takes a simpler form 
\begin{equation}
	\chi(\tau, l)
	=\frac{ka_0}{4}\int^{\tau+l}_{0}d\lambda\;
	H\left(\frac{\lambda-\tau+l}{2}\right)
	\cos \left(k_g\frac{\lambda+\tau-l}{2}\right)+f(\tau-l)
\label{15}\end{equation}
All functions characterizing shape of the waveguide, the function $H(l)$ among them, are periodical with period $L$. Decomposing $H(l)$ in the Fourier series, 
\begin{equation}
	H(l)=H_0+\sum^{\infty}_{n=1} H_n \cos(k_nl+\psi_n), \quad k_n=\frac{2\pi n}{L}, 
\label{16}\end{equation}
we see that the Fourier components give additive contributions into the function $\chi(\tau, l)$ which can easily be calculated with the help of \Eq{15}. The cases $k_n\ne k_g$ and $k_n = k_g$ differ essentially. In the latter case the gravitational-electromagnetic resonance (GER) takes place. 

Let us find contribution of the harmonic which is in resonance with the gravitational wave. Its number is $n=N=k_gL/2\pi$. Its contribution into the phase of the electromagnetic wave is 
\begin{align*}
		\chi(\tau, l)
		& =\frac{ka_0 H_N}{4}\int^{\tau+l}_{0}d\lambda\;
		\cos\left(k_N\frac{\lambda-\tau+l}{2}+\psi_N\right)
		\cos \left(k_g\frac{\lambda+\tau-l}{2}\right)  \\
		& +f(\tau-l)
\end{align*}
Calculating this integral with the relation $k_N=k_g$ taking into account, we have
\begin{align*}
	\chi(\tau,l) 
	& =(\tau+l)\frac{ka_0 H_N}{8} \cos\big(k_g(\tau-l)-\psi_N\big) \\
	& + \frac{ka_0 H_N}{8k_g} \left[\sin \big(k_g(\tau+l)+\psi_N\big)
	-\sin\psi_N\right]
	+f(\tau-l) 
\end{align*}

Let such a wave number $k$ be chosen for the zero approximation in Eq.~(\ref{10}) that $kL$ is multiple of $2\pi$, i.e. the wave number $k$ corresponds to one of the oscillation modes in the absence of the gravitational wave. Then it follows from the condition of periodicity imposed on the phase, 
\begin{equation}
	\varphi(\tau, l+L)=\varphi(\tau, l)+2\pi n,
\label{17}\end{equation}
that the function $\chi(\tau,l)$ should be a periodical function of $l$ with period $L$. This may be provided by the appropriate choice of the function $f(\tau-l)$: 
$$
f(\tau-l) = (\tau-l)\frac{ka_0 H_N}{8}\cos\big(k_g(\tau-l)-\psi_N\big)+f_1(\tau-l).
$$
where $f_1(l)$ is periodic with period $L$. 

Finally we have 
\begin{equation}
	\chi(\tau,l)=\tau\frac{ka_0 H_N}{4}\cos\big(k_g(\tau-l)-\psi_N\big)
\label{18}\end{equation}
where we omitted all terms which remain restricted with $\tau$ infinitely increasing, and took into account only the specific resonance term. 

We see that the phase of electromagnetic wave in any fixed point of the waveguide (at fixed value of $l$) contains, besides the usual linearly increasing term $k\cdot(\tau-l)$, also the term $\chi(\tau,l)$ oscillating with the linearly increasing amplitude. This is the essence of the gravitational-electromagnetic resonance. 

The only one harmonic $H_N$, $N=k_gL/2\pi$, is essential for quantitative characterization of the resonance term in the function $H(l)$. Particularly, if the waveguide is a circle of radius $R$, then function $H(l)$ has a single harmonic with the wave number $k_2=2/R$ and amplitude $H_2=1$. The resonance is achieved in this case at $k_g=2/R$. Therefore, a circular waveguide is a detector tuned on a single frequency of the gravitational wave, while a waveguide of another shape can detect waves of various frequencies according to the spectrum of the function $H(l)$. 

Although for definiteness we talked of a waveguide, in fact the setup may be realized as a system of mirrors forming the shape of the ray of light in the appropriate way. Particularly, the system of two parallel mirrors with the light ray reflecting orthogonally, may be described as a waveguide of the shape of infinitely squeezed ellipse degenerated into a straight-line segment. In this case components $n_1$, $n_2$ of the  tangent to the waveguide change their sign with altering $l$, but their absolute values remain constant. The function $H(l)$ is then constant and the resonance is impossible. 

The above said is valid only in case of a waveguide lying in the plane $x^1$, $x^2$ (in the plane of the front of the gravitational wave). In general case \Eq{14} has to be used instead of \Eq{15}. Consider in this case only the waveguide degenerated into a straight-line segment (the system of two parallel mirrors) and show that it may be in resonance with the gravitational wave provided that the waveguide (the light ray) is not lying in the front of this wave. 

Let the waveguide is degenerate in the straight-line segment not lying in the plane $x^1$, $x^2$ (which corresponds to a pair of parallel mirrors not orthogonal to the front of the gravitational wave). We have then $H(l)=H_0=\mathrm{const}$ and 
\begin{equation}
	x^3(l) = \left\{
	\begin{array}{lll}
	\beta(l-nL) 		\quad & \mbox{for} \quad & nL<l<\left(n+\frac{1}{2}\right)L, \\
	\beta\big((n+1)L-l\big)	\quad & \mbox{for} \quad & \left(n+\frac{1}{2}\right)L<l<(n+1)L
	\end{array}
	\right.
\label{19}\end{equation}
where $\beta=|n_3|=\mathrm{const}$. In this case \Eq{14} takes the form
\begin{align*}
	\chi(\tau, l) 
	& = \frac{ka_0 H_0}{2}  \left[\cos k_g(\tau - l) \int^{l}_{(l-\tau)/2} dl'\, \cos k_g\big(l'-x^3(l')\big)\right.\\
	& +  \left.\sin k_g(\tau - l) \int^{l}_{(l-\tau)/2} dl'\, \sin k_g\big(l'-x^3(l')\big)\right] + f(\tau - l).
\end{align*}

It is clear that a term increasing with time will appear in this expression only if the integrands have non-zero constant components 
\begin{align*}
	A = & \lim_{l\to\infty} \frac{1}{l} \int^{l}_{0}dl'\, \cos k_g\big(l'-x^3(l')\big), \\
	B = & \lim_{l\to\infty} \frac{1}{l} \int^{l}_{0}dl'\, \sin k_g\big(l'-x^3(l')\big).
\end{align*}
The constant components $A$, $B$ of the signal are equal to zero for almost all values of $k_g$ and may be non-zero only if the wave numbers $k_g$ and $k_1 = 2\pi/L$ are in accordance with each other (in resonance), $k_g = k_N = Nk_1=2\pi N/L$. The calculation shows that in this case\footnote{These formulas for the coefficients $A$ and $B$ are new as compared to the original (Russian) text of the paper.} 
$$
	A =  \frac{\beta}{\pi N(1-\beta^2)}\sin\left[\pi N(1-\beta)\right], \quad 
	B =  \frac{\beta}{\pi N(1-\beta^2)}\left\{1 - \cos\left[\pi N(1-\beta)\right]\right\}.
$$
Making use of these formulas and providing, by the choice of the function $f(\tau - l)$, periodicity of the function $\chi(\tau, l)$ in $l$, we have 
\begin{align}
 \chi(\tau, l) &= \frac{1}{4}\tau\cdot ka_0H_0 
	\left[A\cos \frac{2\pi(\tau - l)}{L}
	+ B\sin \frac{2\pi(\tau - l)}{L}\right] \nonumber\\
  &= \tau\cdot\frac{k a_0 H_0 \beta}{2\pi N(1-\beta^2)}\sin\alpha \cdot \cos\left[\frac{2\pi N}{L}(\tau - l) - \alpha\right]
\label{20}\end{align}
where it is denoted 
$$
\alpha = \frac{1}{2} \pi N (1-\beta).
$$ 
In the special case when $\beta = 1$ (the pair of mirrors orthogonal to the direction of the gravitational wave) we have $A=1/2$ and $B=0$. 

Therefore, the gravitational-electromagnetic resonance appears even for a waveguide degenerated into a straight-line segment (the case of a pair of parallel mirrors) provided that this straight-line segment is inclined to the front of the gravitational wave. Ya.B.Zeldovich was the first who noted this (private communication). 

\section{Long waveguide} \label{sec:LongWaveguide}

Consider a long waveguide in the shape of a spiral of radius $R$ and helix pitch distance $h$. The length of a single coil of the helix is equal to 
\begin{equation}
	L=\sqrt{(2\pi R)^2+h^2}.
\label{21}\end{equation}
Those characteristics of the waveguide that will be necessary for the calculation are expressed through this length in the following way: 
\begin{align}
	& H(l) = H_2\cos k_2l, \quad 
	H_2 = -\left(\frac{2\pi R}{L}\right)^2, \quad
	k_2 = \frac{4\pi}{L}, \label{22}\\
	& x^3(l) = n_3 l, \quad n_3 = \frac{h}{L}.
\label{23}\end{align}

Making use of these formulas in \Eq{14}, we obtain the following expression for the correction to the phase of the electromagnetic wave: 
\begin{align*}
			\chi(\tau, l)
		& =\frac{ka_0 H_2}{4}\int^{\tau+l}_{0}d\lambda\;
		\cos k_g\left(\frac{1-n_3}{2}\lambda+\frac{1+n_3}{2}(\tau-l)\right)
		\cos k_2\left(\frac{\lambda}{2}-\frac{\tau-l}{2}\right)  \\
		& +f(\tau-l).
\end{align*}
The resonance takes place if 
\begin{equation}
	k_g(1-n_3) = k_2 = \frac{4\pi}{L}.
\label{24}\end{equation}
Taking into account this relation when evaluating the integral we have 
\begin{align*}
	\chi(\tau, l)
	& = \frac{ka_0H_2}{8} (\tau+l) \cos k_g (\tau-l) \\
	& + \frac{ka_0H_2}{8(1-n_3)k_g} \sin k_g\big[ (1-n_3)(\tau+l) + n_3 (\tau-l)\big] +f(\tau-l).
\end{align*}

The waveguide is not assumed to be closed in this case, therefore, no condition of periodicity has to be imposed on the phase $\varphi(\tau, l)$. Instead, take the initial condition in the form 
\begin{equation}
	\varphi(\tau, 0) = k\tau.
\label{25}\end{equation}
This condition means that the given harmonic signal is applied in the input of the wave-guide at any time. Choosing the arbitrary function $f(\tau-l)$ in such a way that this condition be valid, we obtain
\begin{equation}
	\chi(\tau, l) = \frac{ka_0H_2}{4}\cdot l \cdot \cos k_g(\tau-l)
\label{26}\end{equation}
where the terms remaining restricted at all $l$ are omitted. 

Thus, the phase of the wave at the output of the long spiral waveguide oscillates with an amplitude proportional to the length of the waveguide (provided that the condition of resonance \Eq{24} is fulfilled). 

\section{Some variants of gravitational wave detectors} \label{sec:VariantsGravDetectors}

Let us consider how gravitational-electromagnetic resonance can be used for detecting gravitational waves. The signal at the output of a closed (for example circular) waveguide is characterized by the phase $\varphi(\tau, l)$  determined by \Eqs{10}{18}. The spectrum of such a signal contains, besides the principal harmonic having frequency $\omega = ck$ also sidebands lying around the frequencies $\omega\pm\omega_g$ and having width of the order of $1/\Delta t$ with $\Delta t = \Delta\tau/c$ being time of the observation (which is finite because of damping the wave in the waveguide). Ratio of the amplitudes of the signal (sideband) and the principal harmonic is equal to $\omega\Delta t \, a_0 H_N/8$. The entity $H_N$ is of the order of unity, the amplitude $a_0$ of the gravitational wave is connected with the energy flow $I_g$ carrying by this wave, as follows: 
\begin{equation}
	\frac{a_0}{4} = \frac{1}{\omega_g}\sqrt{\frac{2\pi G I_g}{c^3}}
\label{27}\end{equation}
where $G = 6,7\cdot 10^{-8}$\,cm$^3$/g\, sec$^2$ is Newton gravitational constant. 

Exploiting the ``long waveguide'' has its own specific features. In practice this type of the gravitational-wave detector may be realized as a light-guide composed of a great number of thin light-guide fibers. At the moment there are light-guides providing attenuation of the order of 2-3~db/km in energy. For detecting gravitational waves, such light-guide has to be prepared in the shape of a spiral, and the monochromatic signal should be applied to the input (through the special dielectric coat for matching the waveguide with the empty space). Then the signal with the phase determined by \Eqs{10}{26} will appear at the output. The spectrum of this signal contains, besides the principal harmonic of the frequency $\omega$, also two sidebands (of the null width in the ideal case) corresponding to the frequencies $\omega\pm\omega_g$ and amplitudes (in respect to the amplitude of the principal harmonic) equal to $\pi a_0H_2(l/2\lambda)$ where $\lambda$ is the wave length of the light propagating in the light-guide. 

An advantage of the helix light-guide as compared with the closed one is in that the signal turns out to be stationary so that the measurement may be performed for an indefinitely long time. The length $l$ of the light-guide may be considerable. Besides, a number of such light-guides can be connected through the coherent amplifiers of light. Yet the length $l$ is much less than the corresponding value $c\Delta t$ in case of the circular wave-guide. However, since the wavelength is much less for light than for radio waves, the factor $l/\lambda$ in case of the light-guide turns out comparable with the factor $\omega\Delta t = 2\pi c\Delta t/\lambda$ in case of the circular wave-guide. 

Let us point out at one more way to increase the signal of a closed wave-guide. It is advantageous in this case to make longer time $\Delta t$ of propagating the electromagnetic wave in the wave-guide. If it propagates freely, the time is restricted by inevitable loss of energy that very soon (for some seconds in radio-diapason) reduce the amplitude of the wave practically to null. If the energy is pumped, then this process may determine the shape of the signal, so that it is not determined by the gravitational wave. 

Apparently, escape from this dilemma may be in a coherent way of pumping similar to one exploited in quantum generators. 

Let, instead of a passive closed wave-guide, a maser with the resonator in the form of a closed (circular) wave-guide be used in the construction of the gravitational-wave detector. Then the energy in the waveguide is added by the radiation of molecules of the active medium which in turn is induced by the radiation already existing in the wave-guide. The radiation of the molecules turns out then coherent with the radiation already existing. In such a process the energy of the wave propagating in the waveguide may be increased without change of its shape. The energy is pumped into the waveguide, but the device realizing this pumping does not determine the shape of the resulting wave. The latter is determined by the shape of the wave-guide and by the external gravitational wave. 

Even more advantageous would be to make use of a laser working in continuous regime and having the circular resonator (the system of mirrors determining a circular ray). In this case the factor $c\Delta t/\lambda$ determining the amplitude of the signal increases. However, for such a device to work properly, the system of pumping has to actually not influence the frequencies and phases of the radiation in the resonator. To say more precisely, it is necessary that this influence lead to change of the frequencies less than the deviation of the frequencies under the influence of the gravitational wave. This determines sensitivity of the detector. More detailed analysis is necessary to give quantitative estimate. 

In conclusion, the author would like to express his gratitude to V.B.Braginsky for many helpful discussions and valuable information on experimental techniques.

\end{document}